\documentclass{aa}
\usepackage{epsfig}
\usepackage{rotating}

\newcommand{\lesssim}{\mathrel{\hbox{\rlap{\hbox{\lower4pt\hbox{$\sim$}}}\hbox{$<$}}}}
\newcommand{\gesssim}{\mathrel{\hbox{\rlap{\hbox{\lower4pt\hbox{$\sim$}}}\hbox{$>$}}}}

\newcommand{\bmv}{$B-V$}
\newcommand{\bmo}{$B-V_0$}
\newcommand{\teff}{$T_{\rm eff}$}
\newcommand{\nli}{$\log n\rm{(Li)}$}

\newcommand{\nfe}{$\log n\rm{(Fe)}$}

\begin{document}

\title{Lithium evolution in intermediate age and old open clusters: 
NGC 752 revisited
\thanks{Based on observations collected at Telescopio Nazionale Galileo, 
La Palma, Canary Islands}}

   \subtitle{ }

   \author{P. Sestito\inst{1} \and S. Randich\inst{2} \and
        R. Pallavicini\inst{3} }

   \offprints{P. Sestito, email:sestito@arcetri.astro.it}

\institute{Dipartimento di Astronomia,
Universit\`a di Firenze, Largo E. Fermi 5,
            I-50125 Firenze, Italy
\and
INAF/Osservatorio Astrofisico di Arcetri, Largo E. Fermi 5,
             I-50125 Firenze, Italy
\and
INAF/Osservatorio Astronomico di Palermo, Piazza del
                  Parlamento 1, I-90134 Palermo, Italy}

\titlerunning{Li evolution in NGC~752}
\date{Received Date: Accepted Date}

\abstract{
We present new high resolution spectroscopic observations 
of the intermediate age ($\sim$2 Gyr) open cluster \object{NGC~752}.
We investigate the Li vs. \teff~distribution
and we obtain a new accurate determination of the cluster metallicity.
We compare the results for NGC~752 with other intermediate age and
old clusters spanning the age range from the \object{Hyades} ($\sim$0.6 Gyr) to
\object{NGC~188} ($\sim$6--8 Gyr). We find that NGC~752 has a solar
iron content ([Fe/H]$=+0.01\pm0.04$), at variance with early 
reports of sub--solar metallicity. We find that 
NGC~752 is only slightly more Li depleted than the younger 
Hyades and has a Li pattern almost identical to that observed in
the $\sim$2 Gyr old \object{IC~4651} and \object{NGC~3680}.
As for the latter clusters,
we find 
that NGC~752 is characterized by a tight Li vs. \teff~distribution 
for solar--type stars, with no evidence for a Li spread as large as
the one observed in the solar age solar metallicity \object{M~67}.
We discuss these results in the framework of mixing mechanisms
and Li depletion on the main sequence (MS). We conclude that the 
development of a large scatter in Li abundances in old
open clusters might be 
an exception rather than the rule
(additional observations of old clusters are required), and that 
metallicity variations of the order of $\sim\pm$0.2 dex do not 
affect Li depletion after the age of the Hyades.
}
\maketitle
\section{Introduction}\label{intro}
Lithium is present only in the outermost layers of a star, since its
relatively low burning temperature ($T_{\rm{Li}}=2.5\times10^{6}$
K) allows Li destruction in stellar interiors.
The amount of Li depletion strictly
depends on the mixing mechanisms
which transport material located at 
the bottom of the convective zone (CZ) towards
deeper layers, making Li a good tracer of the physical processes
occurring during the various evolutionary phases of a star.

The current scenario for Li evolution in field stars of the solar neighborhood
and in open clusters is very complex
(see, e.g., the review of Pasquini \cite{P00} and references therein)
and so far the mechanism(s) driving Li depletion on the 
MS has/have not been clearly identified.

Observations of open
clusters show that, at variance with the predictions
of standard models
(including only convection as a mixing process),
Li depletion in solar--type stars is effective during MS evolution (e.g.
Thorburn et al.~\cite{T93}).
In addition,
metallicity seems to have a negligible effect on Li depletion
at least up to the age of the Hyades
(see Jeffries \cite{jeffries} and references therein).
Most surprisingly, the open cluster
M~67 (solar age and solar metallicity) is characterized by a large
dispersion in Li abundances for stars with similar effective temperatures
(Pasquini et al.~\cite{P97}, Jones et al.~\cite{jones99}).
On the contrary,
solar--type stars in the $\sim$2 Gyr clusters NGC~3680 and IC~4651
(Randich et al.~\cite{R00}) and in the
6--8 Gyr old cluster NGC~188 (Randich et al.
\cite{R03}--hereafter R03) show a tight Li--\teff~relationship
with the average Li distribution following that of the Li rich stars (the
``upper envelope'') of M~67. The only cluster with age intermediate
between those of the Hyades and M~67 for which some evidence has been
reported of a possible Li dispersion among solar analogs is NGC~752 
(Hobbs \& Pilachowski \cite{HP86}--hereafter HP86). It is worth 
mentioning that a large Li spread is also observed among field stars:
for example, the Sun has \nli$=1.1$, much lower than the
abundance predicted by standard models. At the same time 
there are stars older than the Sun (e.g. $\beta$ Hyi) 
with a much higher Li content (e.g.
Pasquini et al.~\cite{P94}).

The discrepancies between standard theoretical predictions
 and observational results
clearly suggest that convection
is not the only mixing process driving Li depletion: for this reason,
models including extra--mixing mechanisms have been developed
(see Pinsonneault \cite{pinso} and references
therein), but these non--standard models are still
poorly constrained, since they fail in reproducing quantitatively
the observed features of Li evolution.

In this context, we have carried out
high resolution spectroscopic observations of NGC~752,
an unusually close (distance $\sim$360 pc, Friel \cite{friel};
${m_V-M_V=8.25\pm0.10}$, Daniel et al.~\cite{daniel}--hereafter D94)
2 Gyr old cluster (Friel \cite{friel}),
with estimated sub--solar metallicity ([Fe/H]$=-0.15$, D94);
the reddening towards the cluster is $E(B-V)=0.035$ (D94). 
NGC~752 is one
of the best studied intermediate age open clusters, but surprisingly
the only published Li data for this cluster date back to some 20 years 
ago (HP86; Pilachowski \& Hobbs \cite{PH88}).
Most important,
the sample of HP86 includes very few solar--type stars
(while in Pilachowski \& Hobbs \cite{PH88} only F--type stars
are studied); in spite of the small
sample, as mentioned, HP86 found some evidence 
for the presence of a Li spread for solar analogs,
although this issue was not discussed in their paper.
We observed a much larger number of solar--type stars in
NGC~752 in order to proof or dis--proof the possible
presence of the scatter; moreover,
the same target can be used to investigate
the dependence of Li evolution on the
metallicity by comparing with other $\sim$2 Gyr old clusters:
IC~4651 and NGC~3680, the first with over--solar
iron content ([Fe/H]$\sim+0.10/0.11$,
Pasquini et al.~\cite{P04}, Carretta et al.~\cite{carretta}), the second
with [Fe/H]$=-0.17$ (Pasquini et al.~\cite{P01}).

In this paper we present our new observations of solar--type stars
in NGC~752, extending the previous survey of HP86.
In Sect.~2 the observations and the abundance analysis are described;
in Sects.~3 and 4 we present our results and a discussion, and
finally in Sect.~5 we report our conclusions.
\section{Observations and abundance analysis}\label{obser}

\begin{figure*}[!ht]
%\hspace{-2cm}
\psfig{figure=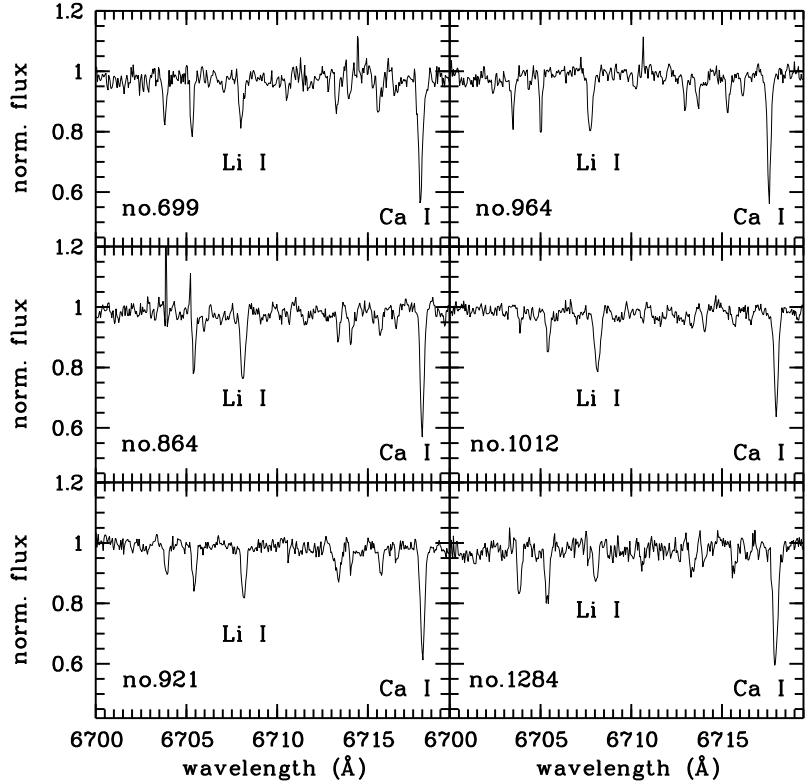, width=15cm, angle=-90}
\caption{Sample spectra in the Li region.}\label{spettri}
\end{figure*}

\subsection{The sample}\label{reduction}
Our sample includes 18 G--type stars selected from the $UBV$
photometry of D94; we selected only stars classified as members 
by D94, whose results are based on previous proper--motion 
and radial velocities studies and on new radial velocities determined
by them (see the discussion in the quoted reference).
\bmv~values range from 0.54 to 0.73 (approximately F8 to G8) and the
visual magnitudes of the observed objects are in the range
$V$$\sim$12.1--13.6. Target stars and photometry are listed in Cols.~1,
2, and 3 of Table~\ref{sample}; we used the identification numbers
of Platais (\cite{pla}) and we adopted the reddening 
quoted by D94, $E(B-V)=0.035$.

The observations were carried out during November 2002 and October
2003 at the Italian National Telescope Galileo (TNG), La Palma
(Canary Islands), equipped with the SARG spectrograph (Gratton et
al.~\cite{gratton}). Target stars were observed using the Red
Cross--Disperser (spectral range 4960--10110 $\rm{\AA}$) and the
OG570 filter, together with the mosaic of two EEV 
CCDs (2048 $\rm{\times}$ 4096; 
pixel size 13.5 $\rm{\times}$ 13.5  $\rm{{\mu}m}^{2}$); we employed a
2$\rm{\times}$2 CCD binning and slit aperture of 0.8'', resulting
in a nominal resolving power $R$$\sim$57,000. The total exposure
times, listed in Col.~4 of Table~\ref{sample} range from 1800 to
7200 sec, providing signal--to--noise ($S/N$) ratios of $\sim$30--80.

Data reduction was performed using the ECHELLE
context within
the package MIDAS and following the usual steps:
first, we separated and flipped the two CCDs, then
we performed bias subtraction, flat fielding, order definition,
order extraction,
sky and scattered light subtraction, and wavelength calibration.
Figure \ref{spettri} shows examples of NGC~752 spectra in the
Li~{\sc i}~${\lambda}6707.8\, {\rm \AA}$ region.

\begin{table*}[!ht] \footnotesize
\caption{Photometry, exposure times, effective temperatures, 
Li equivalent widths and Li abundances
for the target stars. Asterisks denote $EW$s
corrected for the contribution of a nearby iron line (see text).}\label{sample}
\begin{tabular}{cccccccc}
\hline
\hline
no. & $V$ & \bmo & Exposure time & \teff & $EW$(Li)& \nli$_{N\rm{LTE}}$\\
P & & & (s) & (K) & ($\rm{m\AA}$) &\\
\hline
475& 12.847& 0.589 &2200+1800 & 5932 & 59$\pm$10& 2.48${\pm0.16}$\\
520 &12.850& 0.535&3600& 6151& 67$\pm$12 &  2.70${\pm0.16}$ \\
552 &12.921& 0.546&3600+3000& 6106& 61$\pm$7 &   2.62${\pm0.14}$\\
648 &12.108 & 0.550&3600+1800& 6089& 24$\pm$8* &  2.20${\pm0.17}$\\
699 & 13.001 & 0.592&2000+2000 & 5920 &48$\pm$10 & 2.37${\pm0.17}$\\
701 & 13.060 & 0.655&3600 & 5677 &42$\pm$12* & 2.12${\pm0.20}$\\
786 &13.170& 0.695 &3600+1200&  5531 & 20$\pm$7&  1.67${\pm0.23}$ \\
859 &13.200& 0.635 &2700+2700& 5753 & 26$\pm$7& 1.97${\pm0.19}$ \\
864 & 12.885 & 0.543&3600 & 6118 &75$\pm$9 & 2.74${\pm0.15}$\\
889 & 12.802 & 0.516&3600 & 6231 &78$\pm$11 & 2.84${\pm0.15}$\\
921 & 12.644 & 0.518&3600 & 6223 &54$\pm$9 & 2.65${\pm0.15}$\\
964 &12.912&0.547&3600+3600&  6102& 64$\pm$9 &2.64${\pm0.16}$ \\
983 & 13.110 & 0.575 &2000& 5986 &45$\pm$7 & 2.38${\pm0.15}$\\
993 &13.590&0.673&3600+1800& 5611& 20$\pm$6 & 1.73${\pm0.21}$ \\
1012 & 12.417 & 0.504&3000 & 6282 &74$\pm$10 & 2.85${\pm0.15}$\\
1107 &13.660&0.625&1800& 5791& 24$\pm$5 & 1.96${\pm0.17}$ \\
1284 & 12.893 & 0.642&1800 & 5726 &42$\pm$9& 2.16${\pm0.17}$\\
1365 &13.290&0.675&2700+2700& 5603& 48$\pm$8 & 2.11${\pm0.17}$\\
\hline
\end{tabular}
\end{table*}
\begin{table*}[!ht] \footnotesize
\caption{Photometry, effective temperatures, Li equivalent widths and 
Li abundances
for the sample of HP86. Stars marked with asterisks are in common with us.}\label{HP86}
\begin{tabular}{cccccccccc}
\hline
\hline
no. & no. & $V$ & \bmo & \teff &$EW$(Li) & \nli$_{N\rm{LTE}}$\\
P &H & & & (K) & ($\rm{m\AA}$)&\\
\hline
575 & 94 & 13.840 & 0.725 & 5425 & $<$5& $<$0.94\\
701*& 146 & 13.060 & 0.655 & 5677 &  45 & 2.15${\pm0.20}$\\
790 & 185 & 12.267&  0.494 & 6325 &  42 & 2.60${\pm0.19}$\\
859* & 207 & 13.200 & 0.635 & 5753 &  7 & 1.38${\pm0.20}$\\
917 & 227 & 14.100 & 0.895 & 4885 &  $<$12& $<$0.82\\
921* & 229 & 12.644 & 0.518 & 6223 &  63 &2.72${\pm0.19}$ \\
953 & 237 & 12.352 & 0.565 & 6027 &  43& 2.39${\pm0.21}$\\
1017 & 265 & 13.260 & 0.625 & 5791 &  15& 1.75${\pm0.20}$\\
\hline
\end{tabular}
\end{table*}
\subsection{Lithium}\label{lithium}

Li abundance analysis was carried out as in R03; effective temperatures
(listed in Col.~5 of Table~\ref{sample}) were estimated from
dereddened \bmv~colors using the calibration of Soderblom et al.
(\cite{S93}--hereafter S93):
$T_{\rm eff}=1808{(B-V)_{0}}^{2}-6103(B-V)_{0}+8899$ K.
The photometry of D94 is based on 
photoelectric data derived from 
six different sources combined and transformed
into a common $BV$ system: typical errors in \bmv~are
$\sim$0.010--0.020 dex, with some stars
having $\Delta{(B-V)}$ as large as 0.040; for several stars 
only one source is available and the error in \bmv~is not quoted.
In order not to underestimate the uncertainties
in effective temperature and Li abundances, we adopted the conservative value
${\Delta{(B-V)}=\pm0.040}$;
this uncertainty in \bmv~results into an error in \teff~of $\pm$150 K.

Li abundances (\nli$=12+\log N_{Li}/N_{H}$) were determined from 
measured equivalent widths ($EW$, Col.~6) 
of the Li~{\sc i}~${\lambda}6707.8\, {\rm \AA}$
feature and effective temperatures by interpolation of the curves
of growth (COG) of S93. Given the nominal resolving power and the
low rotational velocities of the stars, the Li line should not be
blended with the nearby Fe~{\sc i} feature at ${\lambda}6707.44\,
{\rm \AA}$; however, in the case of stars P648 and P701 (marked
with asterisks in Table~\ref{sample}), due to low $S/N$ ratios we were not able
to separate the Fe~{\sc i} feature from the Li~{\sc i} line,
thus their $EW$s were corrected for the Fe contribution using
the prescriptions of S93: namely,
$EW$(Fe)$=20(B-V)_{0}-3\,{\rm m\AA}$. 
We then applied 
non--local thermodynamic effect ($N$LTE) corrections to LTE
Li abundances using
the code of Carlsson et al.~\cite{carlsson}. 
\nli$_{N\rm{LTE}}$ are listed in Col.~7 of
Table~\ref{sample}; uncertainties in Li abundances were derived by
quadratically adding errors due to uncertainties in $EW$s and 
\teff.

In the following, we will compare our results for NGC~752 with
those of HP86 and with the Li distribution of other clusters of
different ages and/or metallicities (Hyades--Thorburn et al.~\cite{T93};
NGC~3680 and IC~4651--Randich et al.~\cite{R00}; M~67--Jones et
al.~\cite{jones99}). In order to put all the data on
a homogeneous scale, we reanalyzed the published equivalent widths
with the method described above.

Stars selected from the HP86 sample are listed in Table
\ref{HP86}, using the identification numbers of Platais
(\cite{pla}, Col.~1) and Heinemann (\cite{hei26}, Col.~2; the
latter one adopted by HP86); there are three objects in common with
our sample, marked with asterisks in Table \ref{HP86}. Effective
temperatures (Col.~5) were recomputed by us starting from the
photometry of D94 (Cols.~3 and 4). Note that we have selected only
HP86 stars listed as members by D94 and with \teff~cooler than
6500~K, since stars warmer than $\sim$6300--6400~K
should belong to the Li dip, which will not be discussed in this
paper (see HP86; Pilachowski \& Hobbs \cite{PH88}; Balachandran
\cite{bala95}). Col.~6 shows the $EW$s of the 6708 $\rm{\AA}$ feature: HP86
did not take into account the possible presence of the Fe~{\sc i}
contribution, thus we used their published $EW$s without 
performing any correction.
The nominal resolving power of HP86 spectra is 0.2 $\rm{\AA}$ and
the correction should not be strictly necessary;
however, at this resolution there could be a partial
blending between the Li and Fe lines, thus NLTE Li abundances
listed in Col.~7 of Table~\ref{HP86} 
might be slightly higher than the true abundances, especially
for the coolest stars in the sample that are
mostly affected by the Fe contribution (see the discussion in HP86).
HP86 did not publish errors in $EW$s, but they quoted a 40 \% uncertainty
in Li abundance ($N_{\rm Li}/N_{\rm H}$) without considering the error due to
uncertainties in \teff. Errors listed in Col.~7 were computed
by quadratically adding the two contributions.

\subsection{Metallicity}\label{metal}

\begin{table*}[!ht] \footnotesize
\caption{Stellar parameters for NGC~752 stars.}\label{parameters}
\begin{tabular}{cccccccccccc}
\hline
\hline
     & P475 & P520 & P552 & P699 & P864 & P964 & P983 \\
\hline
\teff$\rm{_{phot}}$ &5932 &6151 &6106 &5920 &6118 & 6102&5986 \\
$\xi\rm{_{phot}}$ &1.11 &1.13 &1.17 &1.11 &1.17 &1.17 &1.13 \\
\teff$\rm{_{spec}}$ &5982 &6120 &6130 &5970 &6060 &6102 & 5930\\
$\xi\rm{_{spec}}$ &1.20 &1.30 &1.30 &1.25 &1.18 &1.17 & 0.92\\
\hline
\hline
\end{tabular}
\end{table*}

\begin{table*}[!ht] \footnotesize
\caption{Fe abundances for stars in NGC~752 and for three stars in the Hyades.
The average metallicities and their random errors are also given.}\label{irontable}
\begin{tabular}{cccccccccccc}
\hline
\hline
Fe~{\sc i} line & & & & &\nfe  & & & & \\
    ($\rm{\AA}$) & P475 & P520 & P552 & P699 & P864 & P964 & P983 &\vline & vB21 & vB182 & vB187 \\
\hline
5930.18 &7.54 &7.55 &-- &-- &7.57 &7.53 &7.59&\vline &-- &7.60 & 7.63 \\
5934.65 &7.60 &7.55 &7.62 &-- &-- &7.57 &7.52&\vline &-- &-- & -- \\
5956.69 &7.60 &7.54 &7.63 &7.47 &--  &--  &-- &\vline   &--  &--  &-- \\
5976.77  &7.48 &7.48 &7.48 &7.61 &-- &7.55 &--&\vline &-- &-- &--\\
5984.81 &7.55 &7.44 &7.45 &7.57 &7.44 &7.55 &--&\vline &-- &7.66 &7.72\\
5987.07 &7.63 &-- &7.59 &7.64 &-- &-- &--&\vline &7.72 &7.70 &--\\
6024.06 &7.44 &7.55 &7.58 &7.40 &7.52 &-- &7.51&\vline &-- &-- &--\\
6056.00 &7.45 &7.50 &7.62 &7.57 &7.46 &7.53 & 7.61&\vline&-- &-- &--\\
6078.49 &7.60 &-- &7.62 &7.49 &-- & 7.53&7.65&\vline &7.71 &7.70 &7.70\\
6136.99 &7.46 &7.53 &7.51 &-- &7.46 &7.50 &7.56&\vline &-- &-- &--\\
6157.73&7.57 &-- &7.63 &7.48 &7.47 &7.50 &7.43&\vline &7.63 &7.68 &7.65\\
6187.99&7.54 &7.51 & 7.57&7.51 &7.49 &7.57 &7.58&\vline &7.72 &7.70 &7.68\\
6200.31 &7.55 &7.52 &7.47 &7.44 &7.58 & 7.50&7.63&\vline &7.68 &7.66 &7.69\\
6315.81 &7.55 &7.54 &7.50 &7.56 &7.48 &7.49 &--&\vline &7.62 &-- &7.69\\
6322.69 &7.58 &7.52 &7.50 &7.60 &7.55 &7.58 &7.62&\vline &-- &7.60 &7.72 \\
6330.85 & 7.52&-- &-- &7.42 &7.52 &-- &--&\vline &7.62 & --& 7.61\\
6336.82&7.57 &-- &-- &-- &7.50 &7.51 &7.49&\vline &-- &-- & --\\
6344.15 &7.46 &-- & --&7.49 &7.53 &-- &--&\vline &-- &-- & --\\
6469.19 &-- &7.55 &7.63 &7.56 &-- &7.52 &--&\vline & --& 7.70&-- \\
6498.94 & 7.49&-- &7.52 &7.61 &7.53 &7.51 &7.55&\vline &7.60 &7.61 &-- \\
6574.23 &7.58 &-- &7.59 &-- &7.45 &7.50 &--&\vline &7.60 &7.61 & --\\
6609.11 &7.46 &7.44 &7.49 &7.57 &7.48 &-- &--&\vline & 7.69 &7.70 &7.71 \\
6627.55 &7.56 &-- &-- &7.58 &7.43 &-- &7.59&\vline &7.63 &-- & 7.62\\
6703.57 &-- &-- &7.59 & --&-- &7.65 &7.58&\vline &7.63 &7.67 & 7.64\\
6725.36 &-- &7.49 &-- &-- &7.59 & --&--&\vline& 7.70 & 7.65 & 7.66 \\
6726.66 &-- &7.55 &7.59 &7.57 &7.46 &7.60 &7.59&\vline &7.70 &7.66 &7.68 \\
6750.15 &7.47 &7.45 &7.61 &7.39 & 7.49&7.57 &7.53&\vline &7.58 &-- &7.62 \\
average  &7.53 &7.51 &7.56 &7.53 &7.50 &7.54 &7.56&\vline &7.66 &7.66 & 7.67\\
${\sigma}_{1}$ &$\pm$0.06 &$\pm$0.04 &$\pm$0.06 &$\pm$0.07 &$\pm$0.05 &$\pm$0.04 &$\pm$0.06&\vline &$\pm$0.05 &$\pm$0.04 & $\pm$0.04\\
${\sigma}_{2}$ &$\pm$0.10 &$\pm$0.10 &$\pm$0.10 &$\pm$0.10 &$\pm$0.10 &$\pm$0.10 &$\pm$0.10&\vline &$\pm$0.09 &$\pm$0.09 & $\pm$0.10\\
\hline
\end{tabular}
\end{table*}

The iron abundance of the cluster was derived using stars with
high quality spectra, listed in Tables~\ref{parameters} and ~\ref{irontable}.
The analysis was carried out using MOOG (Sneden \cite{sneden} --
version December 2000) and Kurucz (\cite{kuru}) model atmospheres: 
as a first step, we adjusted $\log gf$ values
(oscillator strength) by performing an inverse abundance analysis
of the Sun. The solar spectrum was obtained at TNG
during the second observing run (October 2003), 
using the same set--up as for our sample stars and
pointing at the Moon; 
we assumed $\rm{\log n(Fe)_{\odot}=7.52}$,
$T_{\rm eff{\odot}}=5770$ K, $\log g_{\odot}=4.44$ and
$\xi\rm{_{\odot}=1.1\,km\,s^{-1}}$. 

In our analysis of NGC~752 we assumed initial effective temperatures derived
as described in Sect.~\ref{lithium}.
For  all our sample stars we adopted the same
surface gravity $\log g=4.5$, while initial microturbulence values
were derived as
$\xi=3.2 \times 10^{-4} (T_{\rm eff}-6390)-1.3(\log g-4.16)+1.7$
(Nissen \cite{nissen},
Boesgaard \& Friel \cite{bf90}). This relationship 
was obtained using a different \teff~scale and different model atmospheres from those used in this paper (see Sect.~\ref{lithium});
however, we note that (i) $\xi$ obtained from the above relationship are only
assumed as initial values for the microturbulence, and (ii)
as we will show below, 
there is a good agreement between the initial values
and those derived by a ``proper'' spectroscopic analysis. 
Conservative random uncertainties for these two parameters are
$\Delta$$\log g=\pm0.3$
dex and $\Delta$$\xi=\pm0.3$
km $\rm{s^{-1}}$.

For each star we measured the $EW$s of several
Fe {\sc i} lines 
in the wavelength range [5900,6800] $\rm{\AA}$ and we computed
Fe abundances (see Table~\ref{irontable}). 
When a trend of~\nfe~vs. $EW$ or $EP$ (excitation potential)
was found, we adjusted the photometric effective temperature
and the initial value of microturbulence
until the trend had disappeared. 
\teff~and $\xi$ derived from the photometry (``phot'') and 
from the Fe analysis (``spec'') are listed in Table~\ref{parameters}:
note that the differences between the two sets of parameters
are within the assumed errors, i.e.
${\Delta{T_{eff}}=\pm150}$ K (see Sect.~\ref{lithium}) 
and $\Delta$$\xi=\pm0.3$
km $\rm{s^{-1}}$. In particular, all the differences
between photometric and spectroscopic \teff~listed in Table~\ref{parameters}
are smaller than 60 K, thus in our spectroscopic analysis we can assume
a typical conservative error $\Delta{T_{\rm eff}}=\pm100$ K.

We did not correct surface gravities,
since Fe~{\sc i} features are only slightly affected by this parameter;
on the other hand,
the dependence on $\log g$ is stronger in the case of
Fe~{\sc ii} lines, which however are very weak in our stars and cannot
be used for an accurate analysis.
In any case, for 
our sample stars which are all MS dwarfs, we do not expect
surface gravities significantly different from the assumed value
$\log g\rm{=4.5\pm0.3}$.

In the last three rows of Table~\ref{irontable} 
the average \nfe~obtained
for each star and the random errors ${\sigma}_{1}$
and ${\sigma}_{2}$ are listed:
${\sigma}_{1}$ is the standard deviation of the mean
iron abundance, representative of errors mainly due to
uncertainties in $EW$s and $\log gf$,
while ${\sigma}_{2}$ is the random error associated to uncertainties
in \teff, $\log g$ and $\xi$. We computed ${\sigma}_{2}$
by varying each parameter at a time and by quadratically adding
the three related errors: we note that for all the stars
${\sigma}_{2}$ is much larger than ${\sigma}_{1}$,
suggesting that the total random error is dominated by uncertainties in stellar parameters.

We checked for the presence of systematic errors in the determination
of \nfe~by performing the same analysis for three stars
of the Hyades: this cluster has a well studied metallicity
([Fe/H]$\rm{=+0.13}$, e.g. Boesgaard \& Budge \cite{bb89}) and can be safely used
to put our results for NGC~752 on a self--consistent scale.
Hyades stars (vB21, vB182 and vB187, listed
in the last three Cols. of Table~\ref{irontable})
were observed by us with UVES at VLT2 
(the description of the data will be reported elsewhere).
These spectra have 
$S/N$$\sim$200 and a resolving power $R$$\sim$40,000, somewhat lower
than SARG spectra: however,
in our analysis we only used Fe features not blended at the UVES 
resolving power.
Stellar parameters for the Hyades objects were derived consistently with
our analysis of NGC~752: we found \teff=5142 K, $\xi\rm{=0.86\,km\,s^{-1}}$ for vB21,
\teff=5079 K, $\xi\rm{=0.84\,km\,s^{-1}}$ for vB182 and
\teff=5339 K, $\xi\rm{=0.92\,km\,s^{-1}}$ for vB187; 
for the three stars we assumed
a surface gravity $\log g=4.5$.

Finally, we computed the weighted mean iron abundance using all the
stars listed in Table~\ref{irontable}.
We found: a) \nfe${=7.53\pm0.04}$ or [Fe/H]${=+0.01\pm0.04}$ for NGC~752,
i.e. the cluster has a nearly solar metallicity; b)
\nfe${=7.66\pm0.06}$,
or [Fe/H]${=+0.14\pm0.06}$ for the three Hyades stars,
very similar to the value
published by Boesgaard \& Budge (\cite{bb89}).
In the computation of the weighted mean, we 
assumed  ${\sigma=\sqrt{{{\sigma}_{1}}^{2}+{{\sigma}_{2}}^{2}}}$ for each star. We note that
the 1$\sigma$--errors on the mean abundances are
much lower than the typical ${\sigma}$ for each star, 
again suggesting that our adopted uncertainties are conservative. 
Systematic errors such as
the model atmospheres and the line list used, together with
possible biases due to different methods of \teff~estimation and
cluster reddening uncertainties, are in principle by far more important.
However, the [Fe/H] obtained for the Hyades, 
which is in very good agreement with previous estimates,
allows us to exclude the presence of large systematic errors.

In Sect.~\ref{intro} we pointed out that D94 assumed a sub--solar
metallicity for the cluster ([Fe/H]$\rm{=-0.15\pm0.05}$),
choosing this value from a ``by eye'' average of various published 
metallicities. The most recent investigations are the spectroscopic
studies of Hobbs \& Thorburn (\cite{HT92}) and
Friel \& Janes (\cite{friel93}). The first authors found
[Fe/H]$\rm{=-0.09\pm0.05}$ from 8 solar--type stars in
the sample of HP86; Friel \& Janes (\cite{friel93})
found [Fe/H]$\rm{=-0.16\pm0.05}$, based on 
a sample of nine low resolution
spectra ($\Delta{\lambda}=4\,{\rm \AA}$) of NGC~752 giant stars.
The result of Hobbs \& Thorburn (\cite{HT92}) is 
more reliable, since it is based on high resolution spectra
($\Delta{\lambda}=0.2\,{\rm \AA}$), but in any case their average
[Fe/H] is rather different from our estimate.

On the other hand, Chaboyer et al.~(\cite{cdp95}) 
adopted a solar iron content for the cluster in a comparison
between the HP86 sample and 
theoretical models including rotation and diffusion.
Our estimate of the metallicity of NGC~752 is perfectly
self--consistent and based on the comparison with the Hyades,
for which we found [Fe/H] in agreement with Boesgaard \& Budge (\cite{bb89}).
Therefore, we can safely assume a nearly solar metallicity for the cluster.

\section{Results}\label{results}

\begin{figure}
\psfig{figure=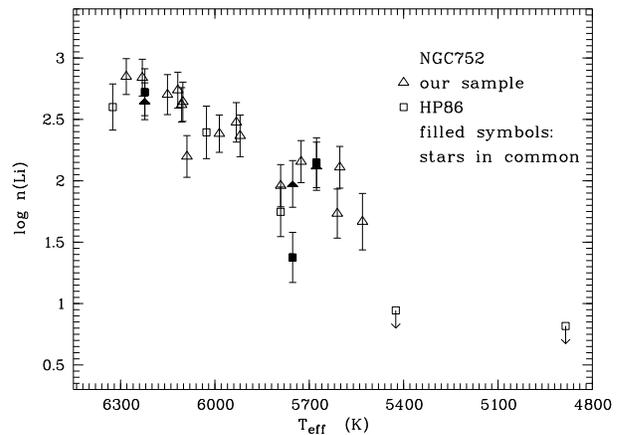, width=6cm, angle=-90}
\caption{\nli~vs. \teff~for NGC~752~-- comparison between
our sample (triangles) and that of HP86
(squares). Filled symbols represent stars in common
between the two samples; down--pointing arrows indicate upper limits
in \nli.}\label{our_hobbs}
\end{figure}

Figure~\ref{our_hobbs} shows the Li distribution (\nli~vs.~\teff)
of NGC~752: our sample is compared to that of HP86.
Li abundances for the three stars in common between the
two samples are plotted as filled symbols: there is a very good
agreement for P701 and P921 (\teff=5677 K and \teff=6223 K,
respectively), while in the case of P859 (\teff=5753 K) there is a large
discrepancy between the two measurements. HP86 reported $EW=7$
$\rm{m\AA}$ for this object, while we measure $\rm{26\pm7m\AA}$;
as mentioned in Sect.~\ref{lithium}, HP86 did not publish
the values of the errors in $EW$s, but they
indicated a mean uncertainty of about 40 \% in $N_{\rm Li}/N_{\rm H}$ (that
we interpreted to be entirely due to the uncertainty in $EW$). 
For star P859, this would translate into an error of $\sim$0.15 dex in \nli,
or $\rm{\sim3\,m\AA}$ in $EW$,
thus it is evident that the two measurements are in complete
disagreement. This discrepancy could be due to the lower
resolution and possibly to a less than optimum $S/N$ ratio of HP86
spectra: for this reason, in the following we will use our
measurements for the stars in common.

The hottest NGC~752 stars (\teff$>$6000 K) are Li rich, since
they have shallow convective envelopes that do not allow a fast
Li destruction: their \nli~values are only
$\sim$2--3 times lower than the meteoritic abundance
(\nli$_{0}=3.1-3.3$, e.g. Mart\'{\i}n \& Magazz\'u
\cite{meteor}), which is thought to be
representative of the initial Li content.
Li abundances decrease towards lower effective temperatures and,
as expected, the coolest objects in the sample
(\teff$<$5500 K) have only upper limits in \nli~since they
are characterized by deep CZs which allow a fast Li depletion to occur.
Solar analogs (5600$\lesssim$\teff$\lesssim$6000) cluster around
a ``plateau'': more in detail,
stars with \teff$\sim$6000 K have average Li abundances
of $\sim$2.4, while stars around $\sim$5700 K have \nli$\sim$2.
The cluster does not appear to be characterized by
a large amount of spread, as discussed below.

\begin{figure}
\psfig{figure=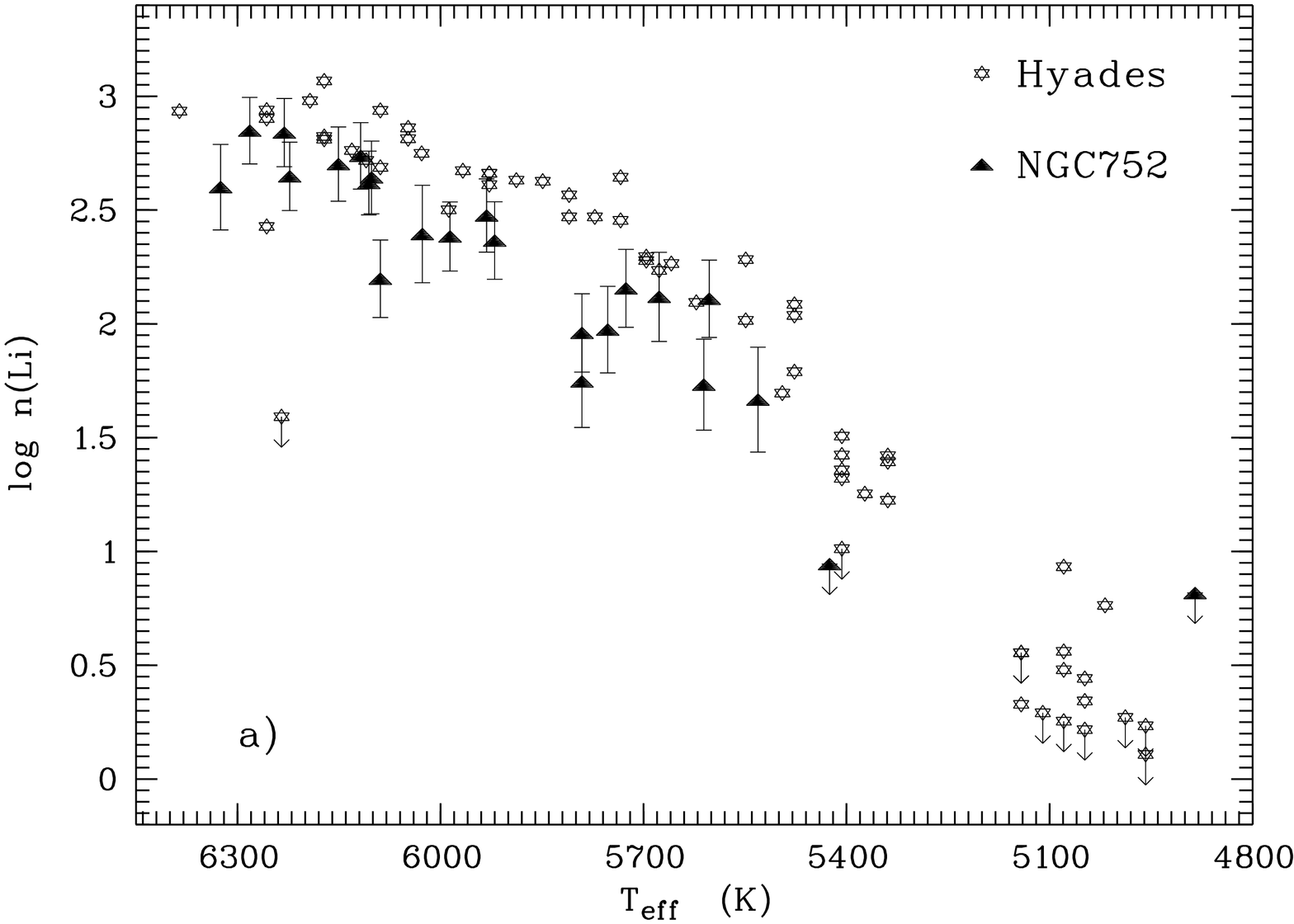, width=6cm, angle=-90}

\psfig{figure=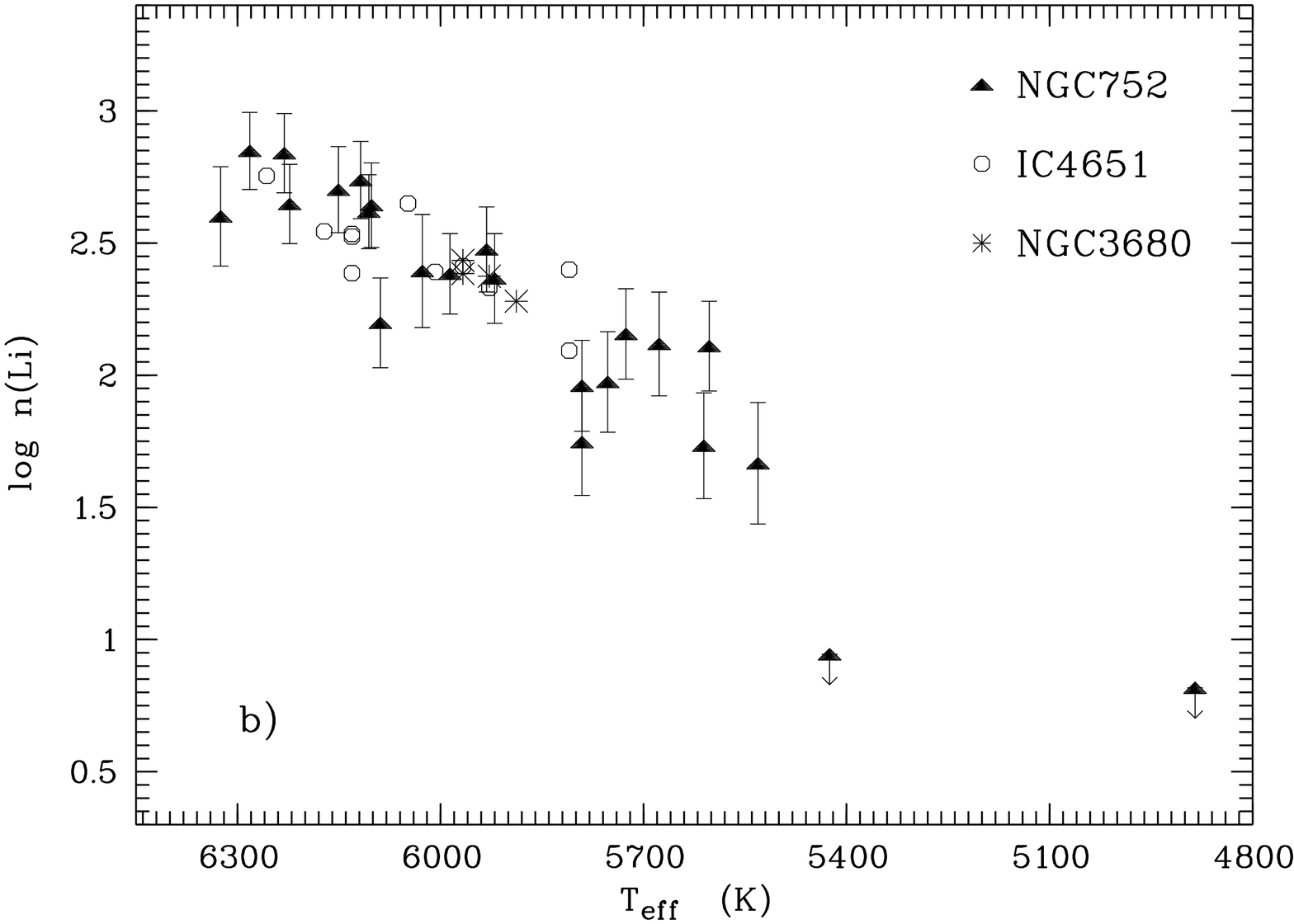, width=6cm, angle=-90}

\psfig{figure=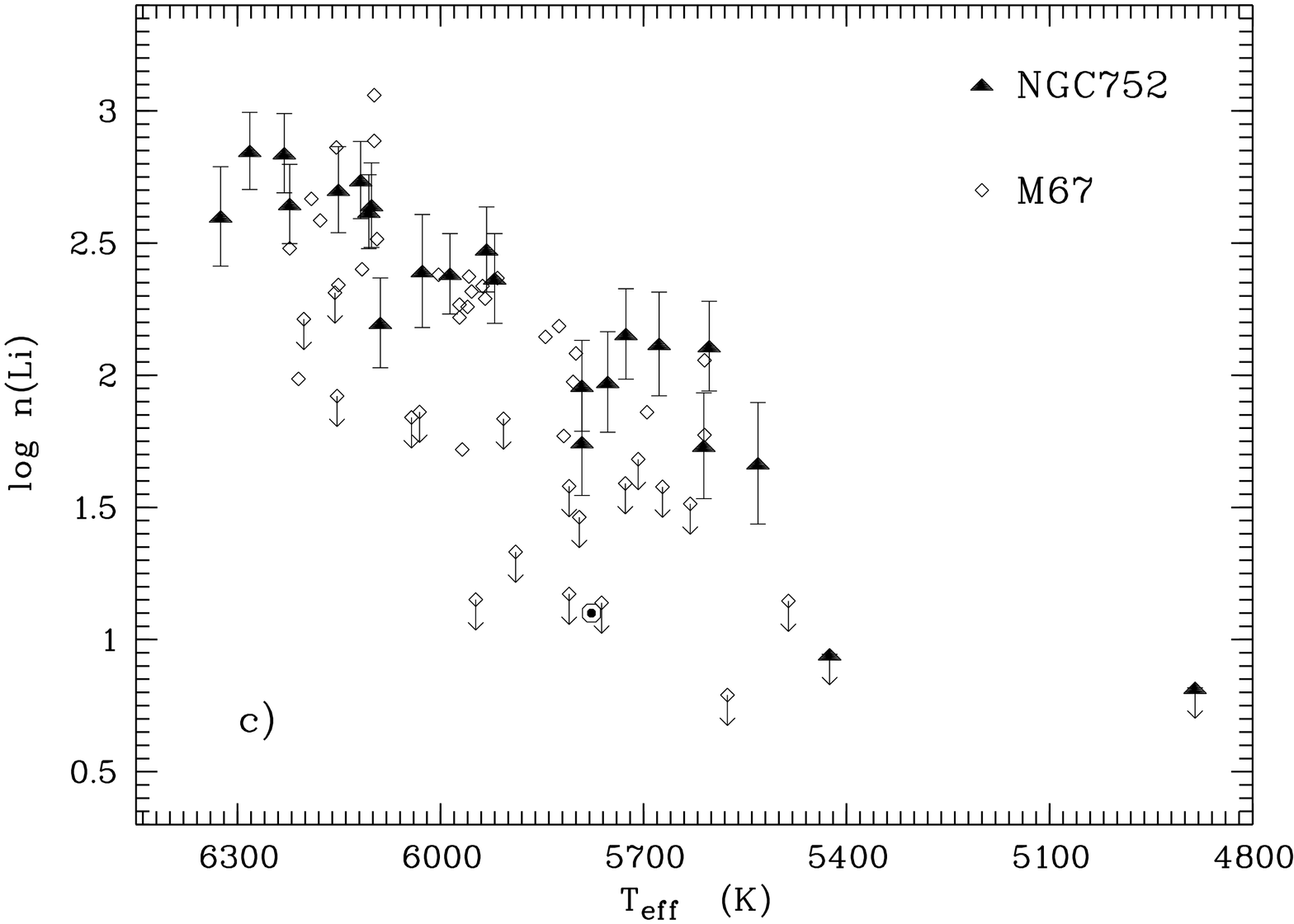, width=6cm, angle=-90}
\caption{\nli~vs. \teff~-- comparison between the
NGC~752 merged sample (our stars+HP86, filled triangles)
and other clusters: {\bf a)} the Hyades (Thorburn et al.~\cite{T93});
{\bf b)} IC~4651 and NGC~3680 (Randich
et al.~\cite{R00}); {\bf c)} M~67 (Jones et al.~\cite{jones99}).}\label{all}
\end{figure}

In Fig.~\ref{all} NGC~752 is compared to various clusters.
In panel a)
we report a comparison between the NGC~752 merged sample
(our stars+HP86, filled triangles) and the 600 Myr old
Hyades (Thorburn et al.~\cite{T93}, starred symbols). 
G--type stars in NGC~752 are somewhat more Li depleted than 
their Hyades counterparts,
but not as much as expected given the difference in age:
\nli~values in NGC~752 are
on average less than a factor 2 lower than those observed in the Hyades,
while the difference between Li abundances in the Hyades 
and those of young clusters (e.g. the Pleiades) is much larger.
Assuming that Li is effectively destroyed during the MS phase,
as shown by observations but at variance
with standard model predictions, the behavior of NGC~752
suggests that Li depletion begins to slow down
after the age of the Hyades (see R03), while
for younger clusters
the rate of Li burning with age for G--type stars is nearly
linear (Sestito et al.~\cite{sestito}).

In Fig.~\ref{all}b NGC~752
is compared to IC~4651 and NGC~3680 (Randich et al.~\cite{R00};
open circles and asterisks):
the three Li distributions are almost indistinguishable.
We remind that 
IC~4651 and NGC~3680 have similar ages ($\sim$2 Gyr) but different
metallicities with respect to NGC~752; therefore, 
age seems
to be the main parameter driving Li depletion during the first 2
Gyr of stellar evolution, while the iron content might
only have second order effects (see Sect.~\ref{iron2}).

Finally, we show a comparison between NGC~752 and 
the solar--age M~67 (Jones et al~\cite{jones99}; 
Fig.~\ref{all}c -- open lozenges).
First of all, we note that the average Li distribution of NGC~752 lies on
the upper envelope of the latter cluster;
second, the most
important feature emerging from this plot is that NGC~752
is characterized by a rather tight Li vs. \teff~distribution,
more similar to that of the Hyades than to the pattern of M~67.

In the latter cluster about 40\% of the stars have severely
destroyed Li, having abundances $\sim$5--10 times 
lower than the remaining ones. On the contrary, in
NGC~752 there are only a few objects slightly deviating from the
average Li distribution
and they should not be considered
as representative of a large Li spread.
Li abundances slightly lower than the general trend could be due
to poor $S/N$ ratios or to possible errors in the photometry.
Therefore, we stress that whereas M~67 is clearly characterized 
by a large Li dispersion, having a considerable fraction of the stars 
that suffered a large Li destruction, NGC~752 has a quite ``regular'' 
Li distribution with all the stars having preserved a relatively 
high Li content.

\section{Discussion}\label{discussion}

In Sect.~\ref{results} we pointed out two main results from our analysis
of NGC~752:
a) the Li vs.~\teff~distribution of the cluster is not characterized
by a significant dispersion;
b) Li depletion seems not to be metal dependent at the age of $\sim$2 Gyr.
\subsection{The star--to--star scatter in Li abundances}\label{scatter}

The first goal of this study was to determine whether a star--to--star 
scatter in Li abundances exists among solar--type stars of NGC~752.
In summary, in the last decade the presence of a considerable
Li spread has been ascertained among 
solar--like field stars (Pasquini et
al.~\cite{P94}) as well as in the solar age, solar metallicity
open cluster M~67 (e.g. Pasquini
et al.~\cite{P97}; Jones et al.~\cite{jones99}).
Randich et al. (\cite{R00}) showed that
no spread is present in the $\sim$2 Gyr old clusters NGC~3680
and IC~4651 and suggested (among
other possibilities) that the dispersion might
develop after this age. 
The more recent investigation
by R03 demonstrated however that also the 6--8 Gyr old
NGC~188 does not show a spread in Li abundances.
NGC~188 is one
of the oldest known open clusters and the oldest open cluster in
which Li has been studied; R03 showed that the
Li distribution of this cluster is similar
to that of NGC~3680 and IC~4651: 
solar--type stars indeed lie on the upper
envelope of M~67, i.e. they are quite Li rich in spite of the old age
of NGC~188. This feature might indicate a slowing down of Li depletion
at ages larger than $\sim$2 Gyr, already discussed in a detailed way 
in the quoted reference.

On the other hand, HP86 data for NGC~752 showed some indication for the
presence of a Li dispersion among solar--type stars, but the sample 
was rather sparse.
As shown by our analysis, based on new high resolution
observations of NGC~752, the cluster is characterized by a tight
Li--\teff~relationship with only a few stars slightly deviating from the
mean trend, which cannot be 
considered as statistically representative of a scatter.
We can demonstrate this by
considering the temperature range 
\teff$\sim$[5500--6300] where M~67 shows
the Li spread (cooler stars have to be excluded since they are
exhausting their Li content, as expected due to the deep CZs):
in this region
we have 20 NGC~752 stars and at most 2 objects are slightly
deviating from the average Li pattern (P648 and P1017), while in the case
of M~67 we conservatively assume that 30 \% of stars are over--depleted.
Thus, if a similar fraction should hold for both the clusters,
using a binomial distribution we have a probability P(2/20)
$\rm{=2.78\times10^{-2}}$
of finding two Li--poor stars in our sample, which is between 
2$\sigma$ and 3$\sigma$.
If we had considered only 1 NGC~752 star as over--depleted and
a more realistic
fraction of Li--poor stars in M~67 equal to 40 \%, we would have found
P(1/20)$\rm{=4.87\times10^{-4}}$ (about 4$\sigma$). 
In both cases we can exclude the 
presence of over--depleted stars in NGC~752; moreover, we stress that
the deviation of stars P648 and P1017 from the
NGC~752 Li distribution is not quantitatively comparable with the spread
observed in M~67: P648 and P1017 are indeed a factor $\sim$2 more
Li depleted than NGC~752 stars of similar temperatures, while stars in
the lower envelope
of M~67 suffered a depletion of factors $\sim$5--10 with respect to their
Li--rich counterparts.

In other words, our analysis points out that (i) M~67
is so far the only old open cluster showing a Li spread
similar to that observed among nearby field stars, i.e.
the development of a dispersion might be
an exception rather than the rule
in intermediate age/old open clusters;
(ii) Li is not 
a good age tracer, at least for open clusters older than $\sim$1 Gyr,
since Li depletion in G--type cluster stars appears to slow down
after the age of the Hyades, but at the same time stars in the field
and in M~67 exist for which Li destruction is unexpectedly very fast.

Several attempts have been made over the last years in order to
understand the possible physical mechanisms responsible for the spread
in M~67 (see the references
quoted below for further details).
For example Jones et al. (\cite{jones99})
proposed the presence of mixing driven by rotation
(models of Pinsonneault et al. \cite{pinso90},
Chaboyer et al. \cite{cdp95}, Mart\'{\i}n \& Claret \cite{martinclaret}),
while Pasquini et al. (\cite{P94}) noticed that also the presence
of mass loss might play a role in the appearing of the scatter
(models of Swenson \& Faulkner \cite{sf92}).
Garc\'{\i}a L\'opez et al. (\cite{GL88})
suggested that M~67 could contain two sub--clusters with different
ages and/or chemical composition. A scatter in heavy element abundances
might indeed affect the tachocline diffusion processes related to light
element burning (Piau et al.~\cite{piau}).

None of the proposed explanation sounds satisfactory, since they
are all in contradiction with the following empirical
evidences: observations of the Li--rotation distribution
in young clusters (e.g. \object{Pleiades} --S93; \object{Alpha Persei} 
--e.g. Randich et al.~\cite{R98}); 
the absence of a Be--Li
correlation (Randich et al.~\cite{R02}) and the absence of a large scatter
in heavy element abundances in M~67 (Randich et al.~\cite{R04}, in preparation).
Moreover, the models of Swenson \& Faulkner (\cite{sf92})
fail in quantitatively reproducing the amount of mass loss suffered by 
the Sun and stars in clusters as young
as the Hyades.

We prefer to attribute the cause of
the Li dispersion observed in M~67 to a possible inhomogeneity
of the cluster, similarly to Garc\'{\i}a L\'opez et al. (\cite{GL88}),
although the origin of this inhomogeneity is not clear and,
as mentioned, it seems not to be related to heavy element abundances.
In any case M~67 appears to be a peculiar cluster,
at least considering the --admittedly small-- dataset for intermediate age/old
open clusters.
In parallel, the problem remains of understanding the similar scatter
observed among solar--like field stars: in particular, the most puzzling
issue is the very poor amount of Li present in the Sun which cannot be
explained by any kind of model.

\subsection{Li depletion and iron content}\label{iron2}

One of the unsolved discrepancies between observations
of Li in open clusters and theoretical predictions
regards the dependence of Li depletion on metallicity.
In fact, as already mentioned, standard models predict
an increase of the opacity and therefore of the depth
of the CZ (meaning a faster Li destruction) as the metallicity
increases.
We mention that also the content of oxygen and $\rm{\alpha}$
elements can affect the opacity (Piau and Turck--Chi\'eze \cite{ptc02}).

On the other hand the empirical picture does not support this point of view,
at least as far as zero--age MS stars in young clusters are concerned.
For example, Jeffries \& James (\cite{blanco1})
carried out a survey of \object{Blanco~1} (100 Myr, [Fe/H]$\rm{=+0.14}$) and they
compared it to the similar age Pleiades ([Fe/H]$\sim$solar), 
showing that the two clusters have similar Li patterns, i.e.
have suffered the same amount of pre--MS Li depletion in spite of the
different iron contents.

Our analysis of Li and Fe abundances in the 2 Gyr old NGC~752,
which we found to have a solar metallicity,
allows us to extend the previous results to older ages.
Figure \ref{all}b shows that G--type stars
in NGC~752
have a Li pattern almost identical to that of stars 
with the same effective temperature
in IC~4651 ([Fe/H]$\rm{=+0.10}$) and 
NGC~3680 ([Fe/H]$\rm{=-0.17}$). The metallicities 
of the three clusters have been derived with different codes/methods, 
and for this reason the comparison could in principle be affected by 
systematic errors; however, we note that the three estimates
are all on the same scale. In fact, 
Pasquini et al. (\cite{P04}), which estimated [Fe/H]$\rm{=+0.10}$ 
for IC~4651 with
the code of Spite (\cite{spite}), obtained the same 
result by carrying out
the analysis with MOOG. Furthermore, they analyzed some
Hyades stars which turned out to have [Fe/H]
in agreement with our estimate and with Boesgaard \& Budge (\cite{bb89}).
Pasquini et al. (\cite{P01}) found instead [Fe/H]$\rm{=+0.06}$
for two Hyades stars, using the code of Carretta \& Gratton
(\cite{cg97}), and for this reason
they scaled by $\sim$0.1 dex the metallicity obtained for NGC~3680,
quoting at the end a best value of [Fe/H]$\rm{=-0.17}$.

In other words, if we assume that the iron contents of the
three clusters were derived correctly and are
on the same scale, as it seems to be, we can say that
variations within $\sim\pm$0.15/0.20 dex in [Fe/H] do not affect
Li depletion, even at ages older than that of the Hyades.

We can also exclude a possible dependence of the Li spread
on the iron content as suggested by the Li patterns of
NGC~752, M~67, and NGC~188 (see
Fig.~\ref{all}c and R03): the three clusters represent 
an age sequence with fixed solar [Fe/H], but only M~67 shows
a dispersion.

\section{Summary and conclusions}\label{concl}

We have presented new high resolution spectroscopic
observations of 18 G--type stars in the $\sim$2 Gyr old open cluster
NGC~752, obtained with SARG at TNG. Our data extend the previous
investigation by Hobbs \& Pilachowski (\cite{HP86}) which was based on
a smaller sample.

We investigated the Li vs.\teff~distribution in this cluster,
as well as the metallicity, and we obtained
the following results:\\
{\it a)} NGC~752 turns out to have a nearly solar iron content 
([Fe/H]${=+0.01\pm0.04}$), at variance with other results in the
literature, which reported a sub--solar metallicity.\\
{\it b)} NGC~752 is only slightly more Li depleted than the younger Hyades
and has a Li pattern almost identical to that observed in
the 2 Gyr clusters IC~4651 and NGC~3680.\\
{\it c)} As the other 2 Gyr clusters
and as the older NGC~188, NGC~752 is characterized
by a rather tight Li distribution for solar--type stars, i.e.
these clusters do not show the large Li spread shown by 
the solar age M~67. Li abundances of NGC~752 stars
are similar to those of the upper envelope (Li rich stars) of M~67.\\

This evidence leads to the conclusions that: \\
{\it (i)} as far as the 
empirical picture is concerned, M~67 
is so far the only cluster
showing a large Li
spread for solar--type stars: as such, it appears to be unusual among
intermediate age and old open clusters. 
This star--to--star scatter is
similar to that observed among nearby field stars, which suggests
that M~67 might represent an inhomogeneous sample. \\
{\it (ii)} since IC~4651 and NGC~3680 have similar ages but different
[Fe/H] with respect to NGC~752 (see Sect.~\ref{iron2}) we conclude that
variations of the metallicity within $\sim\pm$0.2 dex seems to not affect
Li evolution after the age of the Hyades,
extending the results found by other authors
for younger clusters.\\
{\it (iii)} the comparison of the 2 Gyr clusters with the
Hyades, with the upper envelope
of M~67 and with NGC~188 allowed us to confirm that Li destruction
in G--type stars
appears to slow down for very old clusters. This result was already discussed
by Randich et al. (\cite{R03}) and our investigation added a new sample,
providing a higher statistics.\\

Our conclusions are based on a small sample
of old open clusters: three 2 Gyr clusters, the solar age M~67
and the very old NGC~188;
in order to further investigate the above problems, i.e. the dispersion
in M~67, the Li--metallicity relationship, the slowing down of
Li depletion, new observational data are clearly required which would
provide a higher statistical basis.
In this respect, we mention that we have started a project aimed at acquiring Li data
for a large sample of
open clusters older than 1 Gyr.
On the other hand, progress in the theoretical models is also
required in order to understand the more complete observational scenario 
that is coming out from the new observations.

\begin{acknowledgements}
P.S. and S.R. acknowledge support from the Italian Ministero dell'Istruzione,
Universit\`a e Ricerca through a COFIN grant.
P.S. is grateful to the TNG staff for the help during the observations.
We thank the referee, Dr. R.D. Jeffries for his useful 
comments.
\end{acknowledgements}
{}

\end{document}